\documentclass[ aps, twocolumn, prx,citeautoscript,superscriptaddress,floatfix,notitlepage,groupedaddress,nofootinbib]{revtex4-2}
\usepackage[utf8]{inputenc}
\usepackage{natbib}
\usepackage{graphicx}
\usepackage{amsmath}
\usepackage{amssymb,bm,xspace,soul}
\usepackage{color}
\usepackage[dvipsnames]{xcolor}
\usepackage{anyfontsize} 
\usepackage{units}
\usepackage{times}
\usepackage{setspace}
\usepackage{braket}
\usepackage{empheq}
\usepackage{xcolor}
\usepackage[normalem]{ulem}
\usepackage{slashed}
\usepackage{comment}
\usepackage{adjustbox}
\usepackage{wasysym}
\usepackage[bbgreekl]{mathbbol}
\usepackage{float}
\usepackage[caption = false]{subfig}

\DeclareMathAlphabet{\mathcald}{U}{dutchcal}{m}{n}
\SetMathAlphabet{\mathcald}{bold}{U}{dutchcal}{b}{n}
\DeclareMathAlphabet{\mathalt}{U}{dutchcal}{b}{n}

\DeclareFontFamily{U}{BOONDOX-calo}{\skewchar\font=45 }
\DeclareFontShape{U}{BOONDOX-calo}{m}{n}{
  <-> s*[1.05] BOONDOX-r-calo}{}
\DeclareFontShape{U}{BOONDOX-calo}{b}{n}{
  <-> s*[1.05] BOONDOX-b-calo}{}
\DeclareMathAlphabet{\mathcalb}{U}{BOONDOX-calo}{m}{n}
\SetMathAlphabet{\mathcalb}{bold}{U}{BOONDOX-calo}{b}{n}
\DeclareMathAlphabet{\mathbcalbx}{U}{BOONDOX-calo}{b}{n}

\colorlet{linkColour}{magenta}
\colorlet{citeColour}{OliveGreen}
\colorlet{urlColour}{cyan}

\usepackage[colorlinks=true]{hyperref}
\hypersetup{
    unicode=false,          
    pdftoolbar=true,        
    pdfmenubar=true,        
    pdffitwindow=false,     
    pdfstartview={FitH},    
    pdftitle={Emergent QED3},    
    pdfauthor={Hart Goldman},     
    pdfsubject={Subject},   
    pdfcreator={Hart Goldman},   
    pdfproducer={}, 
    pdfkeywords={keyword1} {key2} {key3}, 
    pdfnewwindow=true,      
    colorlinks=true,       
    linkcolor=linkColour, 
    citecolor=citeColour,        
    filecolor=linkColour,      
    urlcolor=urlColour           
}

\allowdisplaybreaks

\newcommand{\sgn}{\operatorname{sgn}}
\newcommand{\Tr}{\operatorname{Tr}}

\newcommand{\hart}[1]{\textcolor{violet}{\it [HG: #1]}}

\begin{document}

\title{Emergent QED$_3$ from half-filled flat Chern bands}
\author{Xue-Yang Song$^*$}
\author{Hart Goldman$^*$}
\author{Liang Fu}
\affiliation{Department of Physics, Massachusetts Institute of Technology, Cambridge, MA 02139}
\date{\today}
\def\thefootnote{*}\footnotetext{These authors contributed equally to the development of this work}\def\thefootnote{\arabic{footnote}}

\begin{abstract}

In recent years, two-dimensional Dirac materials patterned with a superlattice structure 
have emerged as a rich platform for exploring correlated and topological quantum matter. In this work, we propose that by subjecting 
Dirac electrons 
to a periodic magnetic field with triangular lattice symmetry 
it is possible to realize a \emph{quantum critical phase} of $N_f=3$ Dirac fermion species strongly coupled to an emergent gauge field, or 2+1-D quantum electrodynamics (QED$_3$). 
We demonstrate explicitly  that the QED$_3$ phase 
naturally arises from a Dirac composite fermion (CF) picture, where the periodic magnetic field manifests as a periodic CF potential and transforms the CF Fermi surface into gapless Fermi points. We further show that by breaking the particle-hole symmetry of the TI surface -- either by doping or by introducing a periodic electrostatic potential with zero mean -- our quantum critical phase gives way to a sequence of fractional Chern insulator phases. Our theory illustrates the rich menagerie of quantum phases possible around half filling of a flat Chern band. 

\end{abstract}

\maketitle

\setcounter{footnote}{0}

\emph{Introduction.} 
The unprecedented tunability of two-dimensional (2d) van der Waals materials has led to new opportunities to 
explore novel quantum phases of matter. In particular, a plethora of correlated and topological electron states has been found by engineering 2d systems with a superlattice structure.  
Major examples include moir\'{e} potentials established by 
stacking 2d materials with a twist angle between layers or a lattice mismatch~\cite{cao18,Andrei2020,Andrei2021,Kennes2021,Ghiotto2021,Li2021}, spatially varying strains~\cite{tang14, venderbos16}, and buckling in graphene sheets~\cite{andrei20}. The superlattice patterning of these materials commonly provides a spatial modulation at the length scale of $10$nm, which can give rise to flat bands hosting strong interaction effects. 

In the presence of Coulomb interactions,  
partially filled flat Chern bands at odd-denominator filling fractions can support fractional Chern insulators (FCIs)~\cite{Moller2009,Neupert2011,Sheng2011,Regnault2011,Tang2011,Sohal2018,Sohal2020}. 
Like ordinary fractional quantum Hall (FQH) phases, FCIs exhibit a fractionally quantized 
Hall conductivity, topological order, and fractional excitations, but enriched with the symmetry of the periodic lattice. The search for FCIs in 2d materials has garnered much recent experimental attention~\cite{Spanton2017,Xie2021}. 

Less explored is the case of flat Chern bands at half filling. 
Ordinary Landau levels (LLs) in a uniform magnetic field are experimentally observed to host a range of correlated phases at half filling, 
depending on the LL index, $n$. These include the composite Fermi liquid ($n=0$)~\cite{Jiang1989, Halperin1993}, a non-Abelian FQH state 
($n=1$)~\cite{Moore1991,Fradkin1998,Read1999,Stern2018}, and charge ordered states (higher $n$)~\cite{Lilly1999,Du1999,Fradkin1999nematic,Cooper2002,Fradkin2010,You2014}. We are therefore driven to ask: Can topological flat bands beyond Landau levels support new types of quantum phases at half filling? If so, in what material platforms can they be found? 

In this paper, we demonstrate the emergence of an 
exotic  \emph{quantum critical phase} 
in half-filled flat Chern bands, and we propose a realistic material platform for its physical realization.   
This phase can be described in terms of $N_f=3$ species of emergent Dirac fermions strongly interacting through a dynamical U$(1)$ gauge field, or $2+1$-D quantum electrodynamics (QED$_3$). 
As a strongly interacting conformal field theory, many of the universal properties of QED$_3$ are unknown outside of large-$N_f$ limits, and numerical methods~\cite{Karthik2016,Karthik2016a,Chester2016a,Poland2019,Karthik2019,Xu2019,Albayrak2022,He2022} 
have thus far made limited progress, making the possibility of its physical realization all the more tantalizing. 


We show that the quantum critical QED$_3$ state can be realized from a system of two-dimensional Dirac electrons subject to a periodic 
magnetic field with triangular lattice symmetry and two flux quanta per unit cell. 
In the limit of a uniform magnetic field and at charge neutrality, the $n=0$ Landau level is half filled, 
and, in the presence of Coulomb interactions, the ground state is a composite Fermi liquid. When a periodic modulation in the magnetic field is 
introduced, the composite fermions (CFs) experience a periodic electric potential, leading to the formation of CF bands. 
By solving the CF band structure at mean field level, 
we find that massless Dirac cones appear at three $M$-points of the Brillouin zone. For the case of two flux quanta per unit cell, these emergent Dirac cones cross the Fermi level, i.e. the CF Fermi surface is transformed into three CF Fermi points by periodic field modulation.  The low energy physics of the resulting quantum critical state is thus governed by $N_f=3$ QED$_3$. 

Crucial to our construction is the particle-hole symmetry 
of the physical electrons at half filling of the Chern band, which protects the emergent Dirac fermions of our QED$_3$ phase from becoming gapped. 
We also require the electronic Chern band 
to be sufficiently flat: 
A large enough bandwidth 
will instead favor an ordinary Fermi liquid state at half filling. 
These two conditions are 
simultaneously satisfied in two-dimensional Dirac electron systems at charge neutrality, such as topological insulator (TI) surface states, 
subject to a periodic magnetic field.


We further show that when the electronic particle-hole symmetry is broken with a periodic electrostatic potential, 
the QED$_3$ quantum critical phase gives way to FCI phases with topological orders of $\nu=\pm1/3$ Laughlin states, consistent with earlier studies on FCI phase transitions ~\cite{Lee2018}.  
We also find by doping away from half filling a Jain sequence of further FCI phases. Dirac materials in a periodic magnetic field are thus a simple and robust setup 
realizing both QED$_3$ and FCI phases from a half-filled flat Chern band.  


\begin{figure}[t]
  \includegraphics[width=\linewidth]{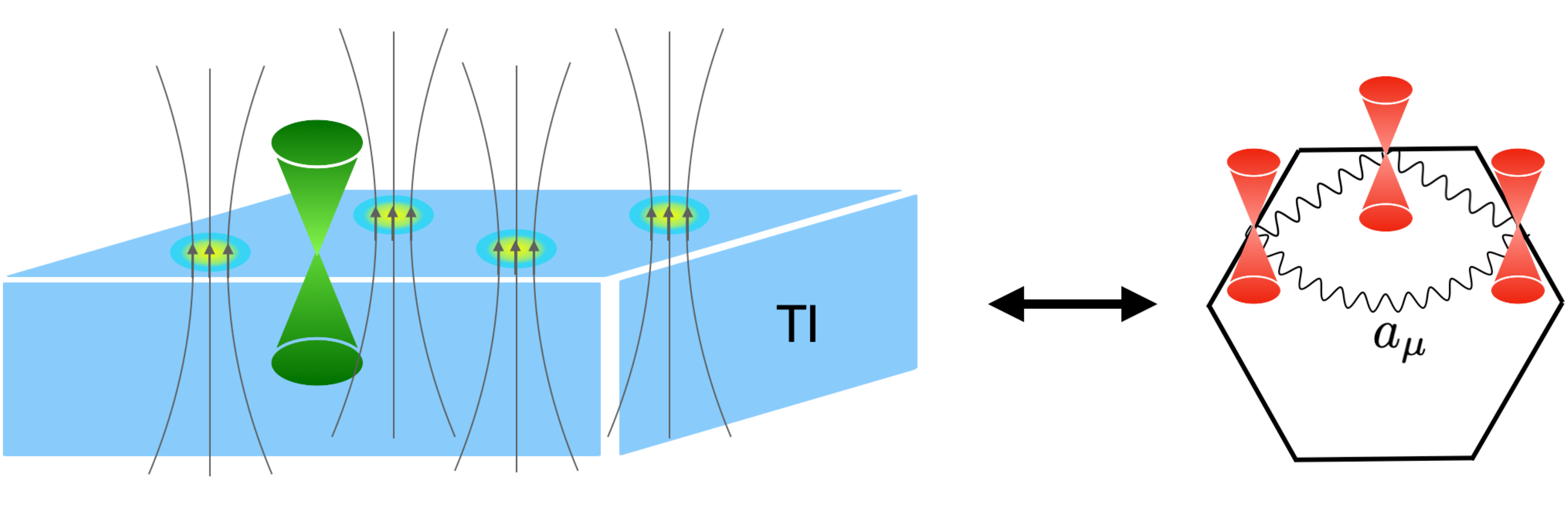}
    \caption{(a) The schematic setup: Topological insulator surface under a periodic magnetic field, forming a triangular vortex lattice in our case, which could arise from proximity to type-II superconductors. The low-energy theory consists of $3$ Dirac fermions interacting with photons (QED$_3$). (b) The Landau level dispersion for free Dirac electrons under a periodic magnetic field where the zeroth Landau level is exactly flat.}
    \label{fig:setup}
\end{figure}

\pagebreak
\emph{Setup.} We begin by considering the TI surface in a uniform magnetic field. The surface of the TI supports a single ``Dirac electron,'' $\Psi$, with action
\begin{align}
S_\Psi=\int_{t,\boldsymbol{x}}
\,i\bar\Psi(\partial_\mu-iA_\mu)\gamma^\mu\Psi\,,
\end{align}
where $A_\mu$ is the background electromagnetic (EM) field, $\gamma^\mu=(\sigma^z,i\sigma^x,i\sigma^y)$ are the Dirac gamma matrices, and we take the Dirac electrons' velocity to be $v=1$. 
Throughout the manuscript, we will use boldface to denote spatial vectors, as well as the notation $\int_{t,\boldsymbol{x}}\equiv \int dt\,d^2\boldsymbol{x}$. 
We also work in units of $\hbar=c=k_B=1$ unless otherwise noted. In the presence of a uniform magnetic field, ${B_0=\boldsymbol{\nabla}\times\boldsymbol{A}_{(0)}}$, $\boldsymbol{A}_{(0)}=\frac{B_0}{2}(y,-x)$, time-reversal symmetry ($\mathcal{T}$) is broken while particle-hole symmetry ($\mathcal{PH}$) ~\cite{Girvin1984} remains intact (see Supplemental Material). Consequently, the Dirac electrons form positive and negative-energy Landau levels. At charge neutrality, the particle-hole symmetry $\mathcal{PH}$ (which exchanges empty and filled states)  guarantees that the $n=0$ Landau level at zero energy is exactly at half filling for any $B_0$. 

It is well-known that in the presence of Coulomb interaction, the half-filled $n=0$ Landau level is a strongly correlated metallic phase known as a composite Fermi liquid (CFL)~\cite{Jiang1989,Halperin1993,Halperin2020}. The existence of the CFL can be explained using flux attachment~\cite{Wilczek-1982,Jain-1989,Zhang-1989,Lopez-1991},  in which each electron is transmuted into a composite fermion via adiabatic attachment of two flux quanta. Attaching flux screens the external magnetic field completely, allowing the composite fermions to form a Fermi surface coupled to a fluctuating U$(1)$ gauge field. Although the traditional flux attachment procedure breaks $\mathcal{PH}$ -- one must decide whether to attach flux to electrons or holes -- Son proposed a new CFL theory~\cite{Son2015} where $\mathcal{PH}$ is manifest, 
\begin{align}
\label{eq:cfaction}
S_\psi=\int_{t,\boldsymbol{x}}\left[i\bar\psi(\partial_\mu-ia_\mu)\gamma^\mu\psi+\frac{1}{4\pi}Ada
\right]\,.
\end{align}
Here the composite fermions, $\psi$, are Dirac fermions 
coupled to a fluctuating U$(1)$ gauge field, $a_\mu$, and we have defined the emergent magnetic field, $b_*=\boldsymbol{\nabla}\times\boldsymbol{a}$\,. From the point of view of the Dirac CFL, the physical electrons in the $n=0$ Landau level are double-vortices of the emergent gauge field~\cite{Wang2015a,MetlitskiVishwanath2016,Seiberg2016,Karch2016,Mross2017,Chen2018,Goldman2018a,Chen2019}, 
\begin{align}
\label{eq:CFeom}
\rho_e=\Psi^\dagger\Psi\leftrightarrow\frac{1}{4\pi}\,b_*\,,\qquad J^i_e=\bar\Psi\gamma^i\Psi\leftrightarrow\frac{1}{4\pi}\,\varepsilon^{ij}\,e_j\,,
\end{align}
where $e_i=\partial_i a_t-\partial_t a_i$. Similarly, like in the well known boson-vortex duality~\cite{Thomas1978,Peskin1977,DasguptaHalperin1981}, the Dirac CFs are vortices of the physical magnetic field: $a_t$ acts as a Lagrange multiplier fixing the CF density, $\rho_\psi=\psi^\dagger\psi$,
\begin{align}
\label{eq: CF density}
\rho_{\psi}=\psi^\dagger\psi=-\frac{1}{2}\frac{B}{2\pi}\,,
\end{align}
meaning that a single CF corresponds to two flux quanta. The 
CFL thus has a circular Fermi surface fixed by the external magnetic field, with Fermi wave vector, ${k_F= \sqrt{|4\pi\,\rho_\psi|} = \sqrt{|B|}}$. 

Importantly, 
the $\mathcal{PH}$ symmetry of the half-filled LL problem appears as a time-reversal symmetry of the Dirac CFs ~\cite{Son2015}, which feel no magnetic field. For clarity, we will denote this symmetry as $\mathcal{T}_{\mathrm{CF}}$ when  discussing the action of physical $\mathcal{PH}$ symmetry on Dirac CFs. 
The detailed symmetry action on the CFs can be found in the Supplemental Material. 

\begin{figure}[t]
 \includegraphics[width=0.75\linewidth]{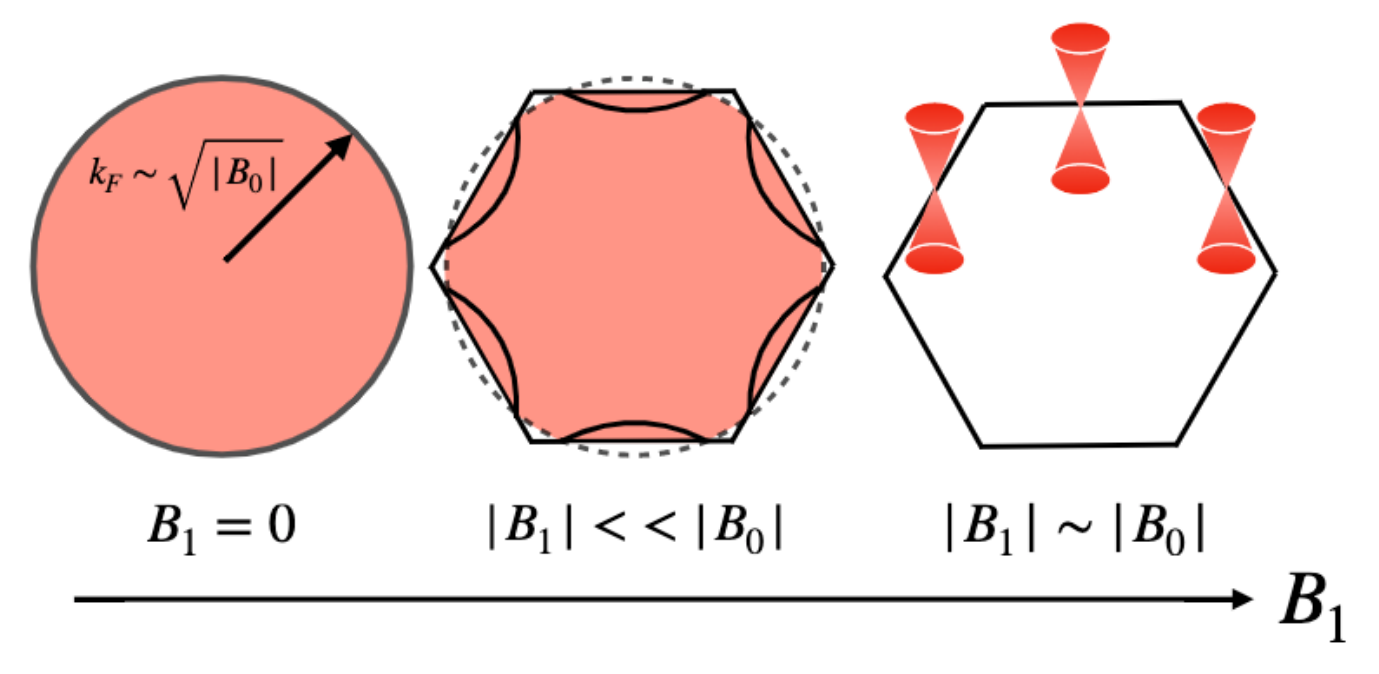}
    \caption{The evolution of the CF Fermi surface at unit filling of the periodic potential. As 
    the strength of the periodic modulation, $|B_1|$ is increased, the CF bands fold at the Brillouin zone edge, and the original circular Fermi surface is deformed. When $|B_1|/|B_0|\sim O(1)$, three isolated Dirac cones emerge at the Fermi energy.} 
    \label{fig:fs_fqh}
\end{figure}

\emph{Emergent $N_f=3$ QED$_3$ from periodic magnetic field.} We proceed to consider the case where the TI surface is subject to a  periodic magnetic field. 
One way such a setup can be established is by placing TI film on top of a type-II superconductor, where an external magnetic field, $H$, induces an Abrikosov vortex lattice, which in turn periodically modulates the magnetic field, $B$, felt by the TI. 
However, in such a setup the flux per unit cell is fixed to the superconducting flux quantum, $h/2e$, rather than the desired case of $2h/e$. 
Alternatively, a periodic $B$-field can be induced by an array of micromagnets \cite{Ye1995,Edmonds2001}. As shown by a recent study \cite{dong22}, in Dirac electron systems under a periodic $B$-field, there remains a perfectly flat Chern band at zero energy, which leads to competing FCI and Wigner crystal states at odd-denominator filling fractions. Our work studies the case of half filling, which corresponds to charge neutrality. It is important to note that the particle-hole symmetry $\mathcal{PH}$ remains exact even when the magnetic field is nonuniform.   

We now consider a magnetic field with both uniform and spatially oscillating components, 
\begin{align}
\label{eq: periodic B}
B(\boldsymbol{x})
&=B_0+2B_1\sum_{n=1}^6\cos\left(\boldsymbol{Q}_n\cdot\boldsymbol{x}\right)\,. 
\end{align}
Here the oscillatory component, which is a two-dimensional periodic function, defines a triangular lattice with lattice constant $\mathfrak{a}$.  
$\boldsymbol{Q}_n=\frac{4\pi}{\sqrt{3}\,\mathfrak{a}}(\sin(\frac{\pi (n-1)}{3}),\cos (\frac{\pi (n-1)}{3}))$, $n=1,\dots,6$, are the reciprocal lattice vectors. 
The CF filling per triangular lattice unit cell is equal to half the number of flux quanta per unit cell, by Eq.~\eqref{eq: CF density}.

In the Dirac CF variables, a slowly varying magnetic field leads to a slowly varying CF density, $\rho_\psi(\boldsymbol{x})=-B(\boldsymbol{x})/4\pi$. Rather than implement this identity as a constraint, it is convenient to instead enforce this relation on average by introducing a scalar potential for the CFs of the same symmetry as the magnetic field $B(\boldsymbol{x})$,
\begin{align}
S_V&=-\int_{t,\boldsymbol{x}}V_{\mathrm{CF}}(\boldsymbol{x})\,\psi^\dagger\psi(t,\boldsymbol{x})\,,\nonumber\\
\label{eq: periodic scalar potential}
V_{\mathrm{CF}}(\boldsymbol{x})&=\mu_0+2V_1\sum_{n=1}^6 \cos\left(\boldsymbol{Q}_n\cdot\boldsymbol{x}\right)\,.
\end{align}
The coefficient, $V_1\propto \ell_{B_0}B_1/4\pi$, $\ell_{B_0}=1/\sqrt{B_0}$, are self-consistently determined by enforcing Eq.~\eqref{eq: CF density} for the mean density, $\langle \rho_\psi\rangle$ (see Supplemental Material)~\footnote{This procedure is valid by the equivalence of thermodynamic ensembles: We have passed from the canonical to the grand canonical formulation of the CF theory.}.

The presence of the periodic potential $V_{\mathrm{CF}}(\boldsymbol{x})$ leads to band folding.    
When the periodic potential in Eq.~\eqref{eq: periodic scalar potential} is made sufficiently strong,  $B_1/B_0\sim\mathcal{O}(1)$, the mean field Dirac CF band structure exhibits three Dirac cones connecting the first and second bands at $E>0$ (see Fig.~\ref{fig:fs_fqh}). 
These Dirac cones are degenerate and located at the $M$-points of the Brillouin zone (BZ), consistent with the $C_6$ lattice rotation symmetry and the CF time-reversal symmetry, $\mathcal{T}_{\mathrm{CF}}$. This CF band structure is depicted in the top panel of  Fig.~\ref{fig:Dispersion}, which is obtained by solving the mean field CF Hamiltonian numerically (see Supplemental Material). Of particular interest to us is the case of unit filling of the triangular lattice, which we denote $f=1$. In this case, there are two flux quanta -- hence a single CF -- at each triangular lattice unit cell. Consequently, the Fermi level of the CFs is exactly at the Dirac points, i.e.  the original CF Fermi surface is transformed into three Fermi points by the periodic modulation of magnetic field.   
Reintroducing gauge fluctuations, the theory then finds itself in an exotic \emph{quantum critical phase} governed by QED$_3$ with $N_f=3$ fermion species.

\begin{figure}[t]
 \includegraphics[width=\linewidth]{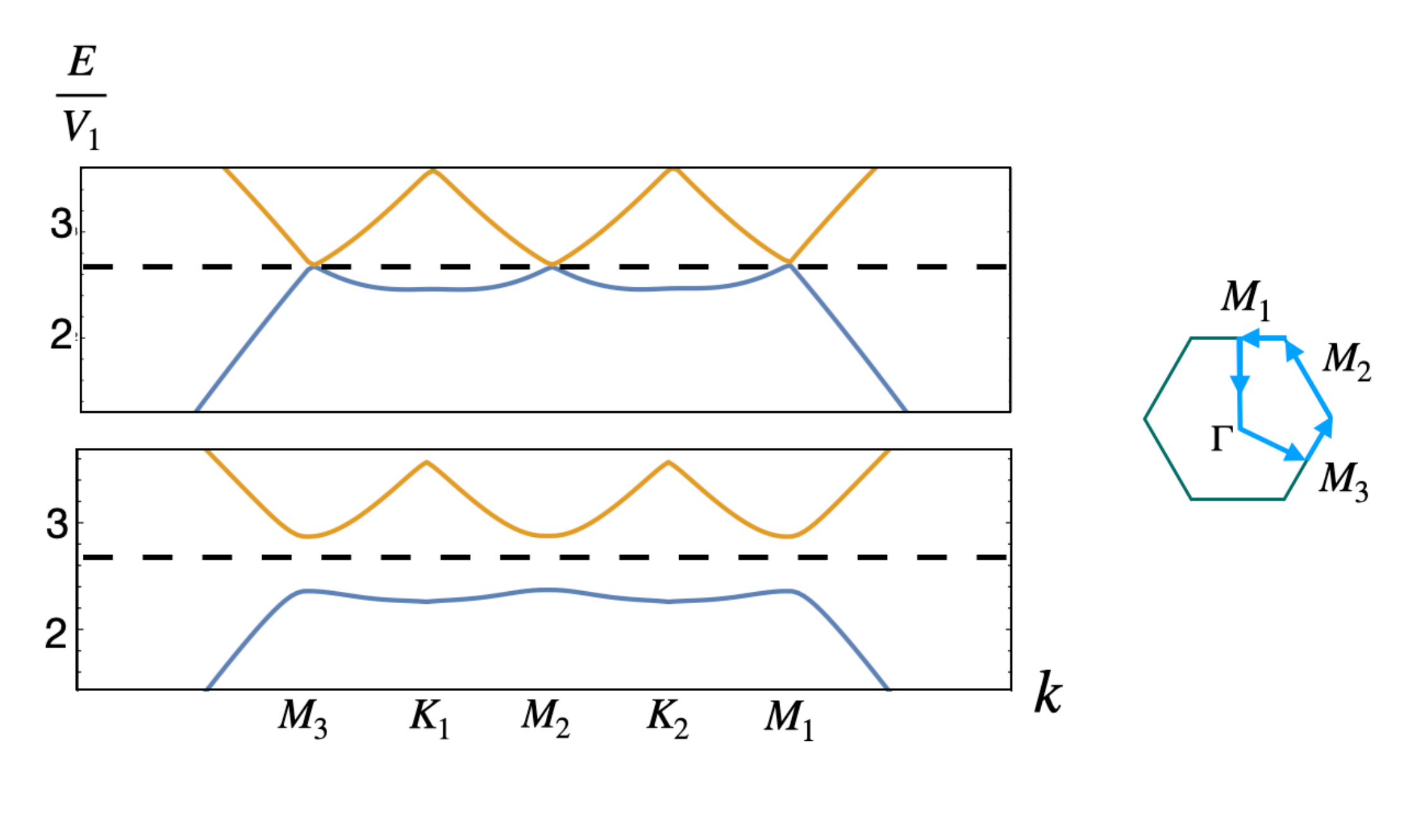}
    \caption{The dispersion of the CFs under a periodic potential, Eq.~\eqref{eq: periodic scalar potential}, with $V_1=2/(\sqrt{3}\mathfrak{a})$. (Top) When the electronic $\mathcal{PH}$ symmetry 
    is enforced, three Dirac cones appear in the CF spectrum at the $M$ points of the BZ. At filling, $f=1$, the chemical potential (dashed line) crosses each of the Dirac points. (Right) There are $3$ Dirac cones along the plotted trajectory at the Fermi energy. 
   (Bottom) Introducing a commensurate, periodic scalar potential for the electrons (magnetic field for the CFs) breaks $\mathcal{PH}$ and causes a gap to open.}
    \label{fig:Dispersion}
\end{figure}

We 
now explicitly calculate the CF dispersion near the BZ $M$-points to leading order in the periodic potential~\cite{wang2021}. 
Without loss of generality, we focus on the points, $\pm \boldsymbol{M}_1=\pm(0,2\pi/(\sqrt{3}\mathfrak{a}))$. The eigenspinors of the mean field CF Dirac Hamiltonian about these points are 
   \begin{align}
   \label{eq: M1 wave functions}
   \xi_+(\boldsymbol{M}_1+\delta\boldsymbol{k})=\frac{1}{\sqrt{2}}\,\left(1, - e^{- i \frac{\sqrt{3}\mathfrak{a}}{2\pi}\delta k_x}\right)\,, \\
   \xi_-(-\boldsymbol{M}_1+\delta\boldsymbol{k})=\frac{1}{\sqrt{2}}\,\left(1,  e^{i \frac{\sqrt{3}\mathfrak{a}}{2\pi}\delta k_x}\right)\,,
   \end{align}
   for small deviations of $\delta\boldsymbol{k}=(\delta k_x,\delta k_y)$ from $\pm \boldsymbol{M}_1$. The periodic 
   potential, Eq.~\eqref{eq: periodic scalar potential}, leads to scattering between the states, 
   $\xi_\pm$, as their momenta differ by $\boldsymbol{Q}_1$.  
   In momentum space, the term which connects these points 
   is $V_1\, \psi^\dagger(\boldsymbol{k})\,\psi(\boldsymbol{k}+\boldsymbol{Q}_1)$.  
   Since the unperturbed eigenspinors of Dirac CFs at $\pm \boldsymbol{M}_1$ form a Kramers pair, the scattering matrix element between them vanishes. In the vicinity of $M_1$-point, the scattering matrix element depends linearly on $\delta\boldsymbol{k}$ as 
   \begin{align}
   V_1 \,\xi_+^\dagger(\delta\boldsymbol{k})\, \xi_-(\delta\boldsymbol{k})=
   i \frac{\sqrt{3}\,V_1\mathfrak{a}}{2\pi}\,\delta k_x\,.
   \end{align}

   The CF dispersion near the $M_1$-point is reconstructed due to scattering off of the periodic potential. 
   If we let $\chi_1(\delta\boldsymbol{k})=c_+(\delta\boldsymbol{k})v_+(\delta\boldsymbol{k})+c_-(\delta\boldsymbol{k})v_-(\delta\boldsymbol{k})$, $v_{\pm}=(\xi_+\pm\xi_-)/\sqrt{2}$, be a generic spinor 
   near the $M_1$-point, we find a Dirac Hamiltonian,
   \begin{align}
   \label{eq:hqed}
   \mathcal H_1=
   \delta k_y\,\chi_1^\dagger\tau^x \chi_1+\frac{\sqrt{3}V_1\mathfrak{a}}{2\pi}\delta k_x\,\chi_1^\dagger \tau^y\chi_1+\dots\,,
    \end{align}  
   where $\tau^{x,y,z}$ are the Pauli matrices in the $v_\pm$ basis. The corresponding results for the Dirac cones near the $M_2$ and $M_3$-points can be obtained using this result by acting with $C_3$ rotations. Notice that the band velocity is anisotropic and thus changes by a $C_3$ rotation between each $M$-point. 

Our mean field result can be substantiated by general considerations based on symmetry. 
Because 
the mean field Hamiltonian involves a single flavor of Dirac CF protected by $\mathcal{T}_{\mathrm{CF}}$ symmetry with $\mathcal{T}^2_{\mathrm{CF}}=-\mathcal{I}$, 
the CF energy spectrum should be gapless, and the total number of degenerate Dirac cones 
must be odd (as in the ordinary case of TI surface states in the presence of time reversal symmetry). 
By Kramers' theorem, the existence of a Dirac point at momentum $\boldsymbol k$ implies the presence of  a degenerate state at $-\boldsymbol k$. This leads to fermion doubling unless  
the Dirac cone appears \emph{only} at BZ points that are left invariant under $\mathcal{T}_{\mathrm{CF}}$, i.e. points satisfying  $\boldsymbol k=-\boldsymbol{k} \mod \boldsymbol{Q}_n$. For the triangular lattice, such points of the BZ are the $\Gamma$-point and the three $M$-points. In particular, 
the $C_3$ symmetry relating the $M$-points implies that Dirac cones at the $M$-points must all be degenerate. 
It is thus natural for the Dirac CF system discussed here to form three Dirac cones at the $M$-points. 

Introducing gauge fluctuations to the mean field result, Eq.~\eqref{eq:hqed}, one obtains at $f=1$ three Dirac fermions coupled to the U$(1)$ gauge field, $a_\mu$. The band velocities of these Dirac fermions are anisotropic, each differing by a $C_3$ rotation. However, the large-$N_f$ QED$_3$ fixed point is stable to both velocity anisotropy and differences in the velocity of each Dirac species~\cite{Hermele2005}. We therefore reasonably expect that the long wavelength theory has emergent Lorentz invariance, and each $\chi_I$ fermion has the same velocity $v\equiv 1$. Hence, $N_f=3$ QED$_3$ arises  as an effective theory at length scales much greater than the period of the oscillatory magnetic field $\mathfrak{a}$, 
\begin{align}
\label{eq: final effective action}
    S_{\mathrm{eff}}&=\int_{t,\boldsymbol{x}} \left[\sum_{I=1}^3 i\bar\chi_I(\partial_\mu-ia_\mu)\gamma^\mu \chi_I+\frac{1}{4\pi}A'da \right]\,,
\end{align}
where again $\gamma^\mu=(\tau^z,i\tau^x,i\tau^y)$, and we define ${A'\equiv A_{\mathrm{total}}-A}$ to be a background probe field on top of the original background field, $A_\mu$, that produces $B(\boldsymbol{x})$. 
Notice that what started as a discrete $C_3$ symmetry in the UV has been enhanced to an emergent SU$(3)$ flavor symmetry, $\chi_I\rightarrow U_{IJ}\chi_J$, at low energies~\footnote{There is a possibility the theory in Eq.~\eqref{eq: final effective action} may be unstable to dynamical mass generation, leading to a trivial state with finite correlation length. See the discussion in Ref.~\cite{Gukov2016} for a review.}. See Supplemental Material for a discussion of how discrete symmetries such as parity and time-reversal act in the effective $N_f=3$ QED$_3$ theory.


Furthermore, our mean field arguments do not depend on the detailed  choice of density-density interaction potential, $V_{\mathrm{int}}(\rho_e)$, for the Dirac electrons, which simply maps to a flux-flux interaction, $V_{\mathrm{int}}(b_*)$, in the dual theory. In particular, we note that instantaneous Coulomb interactions are expected to be marginally irrelevant at the QED$_3$ fixed point~\cite{Lee2020}, while a Lorentz invariant generalization of Coulomb interactions is exactly marginal and leads to a line of fixed points with self-dual properties~\cite{Fradkin1996,Hsiao2017,Mross2017,Goldman2018a}.

\emph{QED$_3$ as a FCI plateau transition.} 
In ordinary quantum Hall systems, incompressible FQH phases are achieved when the filling is tuned away from the CFL state at $\nu=1/2$, with Hall conductivity 
set by the filling. In striking contrast, here we 
 show that incompressible states with fractional Hall conductivity, 
known as fractional Chern insulator (FCI) states, can be induced \emph{fixed at half-filling} for the electronic Chern band, by introducing a periodic chemical potential.

Our QED$_3$ state is protected by the electronic $\mathcal {PH}$ symmetry. When $\mathcal{PH}$ is broken, this critical state can 
become unstable and transition into a new state. 
By introducing a periodic electrostatic potential to break $\mathcal {PH}$, we find the QED$_3$ state gives way to FCI phases, while maintaining half filling of the electronic Chern band. We 
consider a potential that is commensurate with the vortex lattice, 
\begin{align}
\label{eq: periodic mu}
   V_{e}(\boldsymbol{x})=2\mu_1\sum_{n=1}^6 \cos (\boldsymbol{Q}_n \cdot \boldsymbol{x})\,,
   \end{align}
which has zero spatial mean. 
It generates the singlet mass operator, $m\Delta_{\mathrm{singlet}}=m\sum_I\bar\chi_I\chi_I$, with ${\sgn(m)=\sgn(\mu_1)}$. 

From the fermion-vortex  duality, Eq.~\eqref{eq:CFeom}, we see that $V_{e}(\boldsymbol{x})$ sources a periodic magnetic field felt by the CFs, $\overline{b_*}(\boldsymbol{x})$, which we choose to write in terms of a vector potential, ${\overline{a}_i(\boldsymbol{x})=\varepsilon_{ij}\partial_j\overline{\phi}(\boldsymbol{x})}$, where $\overline{\phi}(\boldsymbol{x})$ is a bounded, real-valued function with the same periodicity as $\overline{a}$. 
We find numerically in the lower panel of Fig.~\ref{fig:Dispersion} that 
the combination of periodic potential and periodic gauge field opens a gap in the mean field CF dispersion. In other words, the emergent Dirac fermion in our QED$_3$ state now acquires a finite mass. 

The origin of gap opening can also be understood by a perturbative analysis. 
We proceed analogously to the arguments leading to Eq.~\eqref{eq:hqed}. For momenta near $\boldsymbol{M}_1$, the relevant scattering process induced by the periodic gauge field has wave vectors $\boldsymbol{Q}_{1}$, $\boldsymbol{Q}_4=-\boldsymbol{Q}_1$. 
Denoting the amplitude of $\overline\phi(\boldsymbol{x})$ by $\Phi$ and using the matrix element $\xi_+(0)\gamma^t\gamma^x\xi_-(0)=i$, we find 
that the mean field Hamiltonian contains a Dirac mass term, 
    \begin{align}
    \mathcal H_{1}'
    &=\overline{a}_x(\boldsymbol{Q}_1)\,\psi^\dagger(\boldsymbol{M}_1)\gamma^x\psi(-\boldsymbol{M}_1)+\mathrm{h.c.}\\
    &\propto\frac{\pi\Phi}{\mathfrak{a}}\,\chi_1^\dagger\, \tau^z \chi_1\,,
    \end{align}
where $\chi_1$ was defined above Eq.~\eqref{eq:hqed} and $\sgn(\Phi)=\sgn(\mu_1)$. The analogous calculation can be readily performed for the other two species, yielding a mass, $m\sum_I\bar\chi_I\chi_I$, ${\sgn(m)=\sgn(\mu_1)}$.

When the $\chi_I$ fermions become massive, one obtains a CF Chern insulator with Chern number $3\sgn(m)/2$. Integrating out the fermions and gauge fluctuations, we see that the periodic scalar potential in Eq.~\eqref{eq: periodic mu} tunes an FCI transition across which the Hall conductivity changes by 
\begin{align}
\Delta\sigma_{xy}&=\frac{1}{3}\frac{e^2}{h}\,,
\end{align}
with the two topological orders on each side of the transition corresponding to the $\nu=\pm 1/3$ Laughlin states (see Supplemental Material). 

\begin{figure}[t]

\subfloat{\includegraphics[width=0.5\linewidth]{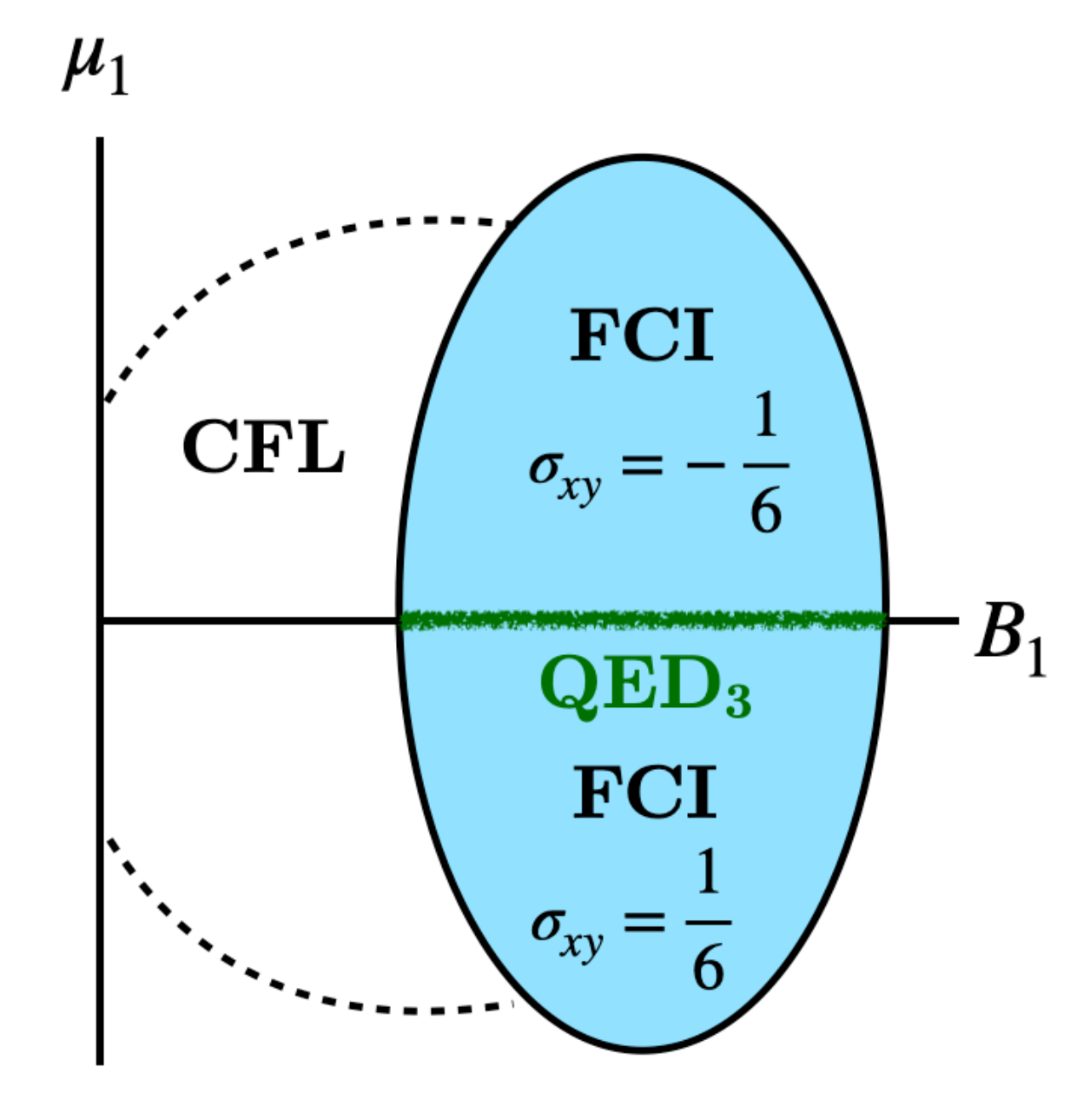}}
\subfloat{\includegraphics[width=0.4\linewidth]{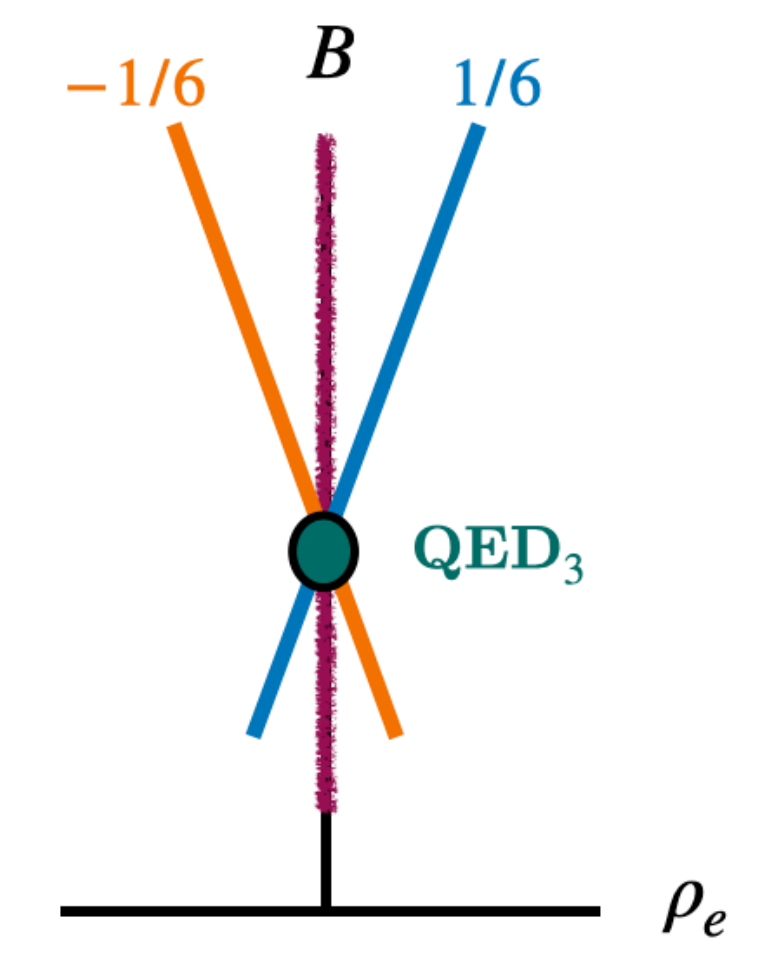}}

  \caption{Two ways to access FCI phases. (Left) By tuning a periodic scalar potential with strength $\mu_1$ while remaining fixed at half filling, the emergent Dirac fermions become massive, leading to FCI states. Note that in this figure the Hall conductivity is that which would be measured on a single TI surface. 
  (Right) By doping away from 
  $N_f=3$ QED$_3$ with a uniform electron density, $\rho_e$, and magnetic field $B'=B-B_0$. In this case, the emergent $N_f=3$ QED$_3$ phase gives way to a sequence of incompressible FCI states. The magenta segment on the $B$-axis denotes the CFL phase. The green dot corresponds to the case of two flux quanta per unit cell, where the quantum critical QED$_3$ phase appears. }
  \label{fig:phase-diagram}
\end{figure}

It is also of interest to consider scalar potentials that break the $C_6$ lattice symmetry by having different amplitudes, $\mu_1^{(n)}$, for each $\boldsymbol{Q}_n$ vector. Such deformations generically introduce an octet of mass operators of the form, ${\Delta^b_{\mathrm{octet}}=\sum_{I,J}\,\bar\chi_I\,t^{b}_{IJ}\,\chi_J}$, where $t^b$, $b=1,\dots,8$, are the generators of $\mathrm{SU}(3)$. These operators can tune the theory to (1) the FCI states described above, as well as (2) a Chern insulator state with integer Hall response occurring when one of the $\chi_I$ fermions receives a mass with opposite sign from the other two, and finally (3) a $\mathcal{PH}$-preserving state where one of the emergent Dirac fermions is gapless and the remaining two are gapped with opposite sign. 

 






\emph{Doping QED$_3$: FCI Jain sequence.} 
We now consider finite doping from charge neutrality --- which breaks the electronic $\mathcal{PH}$ --- and varying $B_0$ away from the case of two flux per unit cell. We find that our emergent QED$_3$ theory gives way to  a family of  FCI states, in analogy with how the CFL is a parent state for the celebrated Jain sequence of observed Abelian FQH phases. 
At finite doping $\rho_e$, the $\chi_I$ fermions feel a uniform emergent magnetic field, $b_*$, and form their own Landau levels (including the zeroth Landau level). When an integer number of these Landau levels are completely filled, an incompressible FCI appears. This occurs when the $\chi_I$ fermions are at filling,
\begin{align}
\label{eq: emergent filling}
\nu_{\chi}=2\pi\sum_I\frac{ \langle  \chi_I^\dagger\chi_I\rangle}{b_*}=3\left(q-\frac{1}{2}\right)\,,\qquad q\in\mathbb{Z}\,.
\end{align}
Here the $1/2$ term comes from the zeroth Landau level of the emergent Dirac fermions in our QED$_3$ state, and the factor of $3$ accounts for the three flavors. From Eq.~\eqref{eq: final effective action}, we observe that the emergent Dirac fermion density is set by the deviation from $B_0$, denoted $B'= \varepsilon_{ij}\partial_i A'_j=- (4\pi)\sum_I\chi^\dagger_I\chi_I$, and that the background charge density leads to an emergent magnetic field, $\rho_e=b_*/4\pi$. 
Combining these relations with Eq.~\eqref{eq: emergent filling}, we see by the Streda formula that the resulting incompressible FCI phase has Hall conductivity, 
\begin{align}
\label{eq: FCI Jain}
    \sigma_{xy}=2\pi\frac{d\rho_e}{dB'}=-\frac{1}{12q-6}\frac{e^2}{h}.
\end{align} 
Here $\sigma_{xy}$ is the Hall conductivity associated with a single TI surface, of which the contribution from the half-filled Chern band is $\sigma^0_{xy}  = \sigma_{xy} + 1/2$ in units of $e^2/h$ (because Chern bands at negative energies contribute $-1/2$).  Comparing $\sigma^0_{xy}=(3q-2)/(6q-3)$
with the Jain sequence, $\sigma^{\mathrm{Jain}}_{xy}=p/(2p+1)$, we find that each of these fractions lies on the principal Jain sequence, 
and has the same topological order as their associated Jain states.  
  Importantly, only the Jain sequence states satisfying $p=3q-2$ appear on the FCI sequence in Eq.~\eqref{eq: FCI Jain}. One therefore expects to measure a Landau fan of FCI states in a Chern band near half filling, with 
  slopes given by Eq.~\eqref{eq: FCI Jain}, as shown in  Fig.~\ref{fig:phase-diagram}. 
  
\emph{Universal properties of the emergent QED$_3$ state.} Among the most basic universal properties of the $N_f=3$ QED$_3$ theory in Eq.~\eqref{eq: final effective action} are the scaling dimensions of the mass operators, denoted $[\Delta_{\mathrm{singlet}}]$ and $[\Delta_{\mathrm{octet}}]$, which determine the divergence of the correlation length as the transition is crossed. For example, tuning the FCI transition with a periodic scalar potential will lead to a diverging correlation length exponent, $\xi\sim (\mu_1)^{-\nu}$, with ${\nu=1/(3-[\Delta_{\mathrm{singlet}}])}$. The values of $[\Delta_{\mathrm{singlet}}]$ and $[\Delta_{\mathrm{octet}}]$  can be calculated in the large-$N_f$ expansion~\cite{ChenFisherWu1993,Rantner2002,Hermele2005,Chester2016}, but the validity of extending these results to small $N_f$ is unclear. Were one to establish an experimental or numerical realization of our setup, it could be possible to measure these exponents, either directly from the correlation functions of the mass operators in numerics, or by measuring the scaling of the DC conductivity, $\sigma_{xx}\sim f(\omega/T,\mu_1/T^{\nu z})$ with $z=1$~\cite{Sondhi1997}. 

The DC conductivity at quantum phase transitions in 2d is another essential piece of universal data. Because the EM current of the theory in Eq.~\eqref{eq: final effective action} is the emergent electric field, $J_i=\varepsilon_{ij}e_j/4\pi$, $e_i=\partial_ia_t-\partial_t a_i$, one obtains for a rotationally invariant system, 
\begin{align}
\sigma\left(\omega\over T\right)=\frac{i\omega}{(4\pi)^2}\,\Pi^{-1}\left(\omega \over T\right)\,,
\end{align}
where $\Pi_{xx}=\Pi_{yy}\equiv\Pi$ is the polarization of the emergent gauge field. Both the DC ($\omega/T\rightarrow 0$) and optical ($T/\omega\rightarrow 0$) conductivities are universal numbers of $\mathcal{O}(e^2/h)$, but they do not necessarily take the same value~\cite{damlesachdev97}. They can again be calculated in the large-$N_f$ limit, which gives~\cite{Goldman2017} 
\begin{align}
\sigma\left({\omega\over T} \rightarrow0\right)\rightarrow\infty\,,\, \sigma\left({T\over \omega}\rightarrow0\right)=\frac{2}{\pi N_f}\frac{e^2}{h}\,,
\end{align}
suggesting that in the clean limit the theory may be a perfect conductor with diverging DC conductivity and vanishing density of states. Note that to obtain this result, one computes \emph{the resistivity} in the large-$N_f$ limit, using ${\Pi(\omega,\boldsymbol{q}=0)=iN_f\omega(\frac{1}{16}+\mathcal{O}(N_f^{-1}))}$ (for $T=0$), and then inverts the result to obtain the conductivity. This is why the leading contribution to the conductivity goes as $1/N_f$. 

Similarly, the electronic compressibility is given by the correlator of the emergent magnetic field, since $\rho_e=b_*/4\pi$. Because of the scale invariance of $N_f=3$ QED$_3$, the static compressibility should vanish linearly as $T\rightarrow 0$, $\kappa(T)\sim T$. This is in contrast to the CFL phase, which has finite compressibility as $T\rightarrow0$.

We finally comment that truly realistic systems have quenched disorder, which will cause QED$_3$ theories to run to a new fixed point governed by an interplay of disorder and interactions~\cite{Goswami2017,Thomson2017,Goldman2017,Goldman2020,Lee2020}. In the $N_f=3$ theory, $\mathcal{PH}$-symmetric disorder (a random magnetic field component for the electrons) is exactly marginal, leading to a line of fixed points with varying dynamical critical exponent, $z>1$. For generic disorder breaking $\mathcal{PH}$, the theory runs to strong disorder and gives way to diffusion ($z=2$), leading to a quantum critical point governed by a gauged non-linear sigma model (NLSM)~\cite{Kumar2019,Lee2021} which at mean field level resembles Pruisken's theory of the IQH transition~\cite{Pruisken1984,Pruisken1985}. Determining the universal properties of NLSMs of this type remains an important open problem.

\emph{Discussion.} In this work, we have demonstrated that half-filled Landau levels of Dirac electrons in spatially periodic magnetic fields can give rise to exotic quantum critical states -- namely, $N_f=3$ QED$_3$ -- without any fine tuning. 
This critical state is protected by the particle-hole symmetry of Dirac electron systems (such as TI surface or graphene) at charge neutrality. 
More generally, an emergent particle-hole symmetry may be found  in 
other Chern band systems at half filling, as it is the case in half-filled Landau levels at high magnetic fields.  
Thus, the exotic quantum critical physics described in this work may be a more general feature of half-filled Chern bands. Speaking more broadly, half-filled Chern bands provide ample opportunities for realizing novel quantum states of matter and call for 
much further study.

\emph{Acknowledgements.} We thank P.A. Lee, T. Senthil, A. Vishwanath, and J. Wang for discussions. HG is especially grateful to R. Sohal for enlightening conversations. This work is funded by the Air Force Office of Scientific Research (AFOSR) under award FA9550-22-1-0432. HG and XYS were supported by the Gordon and Betty Moore Foundation EPiQS Initiative through Grant No.~GBMF8684 at the Massachusetts Institute of Technology. LF is partly supported by the David and Lucile Packard Foundation. 

\bibliography{QED3}
\clearpage

\onecolumngrid

\section*{Supplemental Material}

\subsection{Mean field CF band structure}

Here we describe the numerical calculation of the mean-field CF band structure in the presence of a a periodic modulation, Eq.~\eqref{eq: periodic B}. After mapping to the Dirac CF picture, one has a Dirac cone lying at the $\Gamma$ point below the Fermi energy. As described in the main text, the periodic magnetic field can be viewed as a periodic CF chemical potential. Thus, 
at leading order, the periodic magnetic field scatters the CFs with a momentum difference of the reciprocal super-lattice vectors, $\boldsymbol{Q}_n$. In momentum space, the single particle Hamiltonian is
\begin{eqnarray}
\mathcal H_{CF}=\int d^2 \boldsymbol{p} \,\bar\psi(\boldsymbol{p})\,(\gamma^i p_i)\,\psi(\boldsymbol{p})+V_1\sum_{n=1}^6\psi^\dagger(\boldsymbol{p})\psi(\boldsymbol{p}+\boldsymbol{Q}_n).
\end{eqnarray}

\begin{figure}[t]
    \centering
             \adjustbox{trim={.35\width} {.01\height} {.0\width} {.01\height},clip}
  { \includegraphics[width=0.75\linewidth]{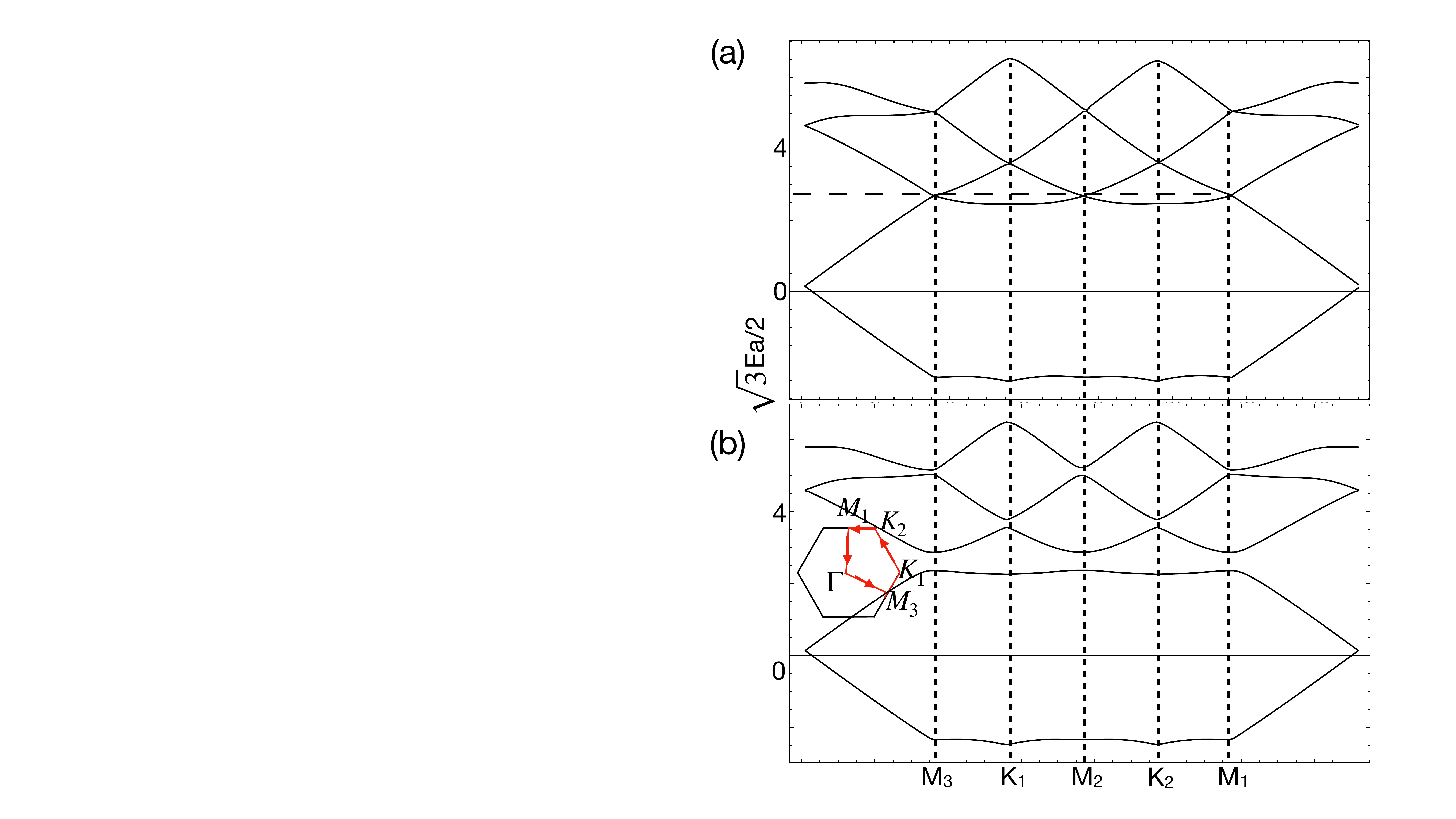}}
    \caption{The dispersion of the CFs under a periodic potential, Eq.~\eqref{eq: periodic scalar potential}, with $V_1=2/(\sqrt{3}\mathfrak{a})$ showing more bands and Dirac fermions at $M,K$ points. Physical settings are identical to main text Fig \ref{fig:Dispersion}. }
    \label{fig:supp_dispersion}
\end{figure}

\begin{figure}[t]
    \centering
             \adjustbox{trim={.01\width} {.15\height} {.01\width} {.05\height},clip}
  { \includegraphics[width=0.5\linewidth]{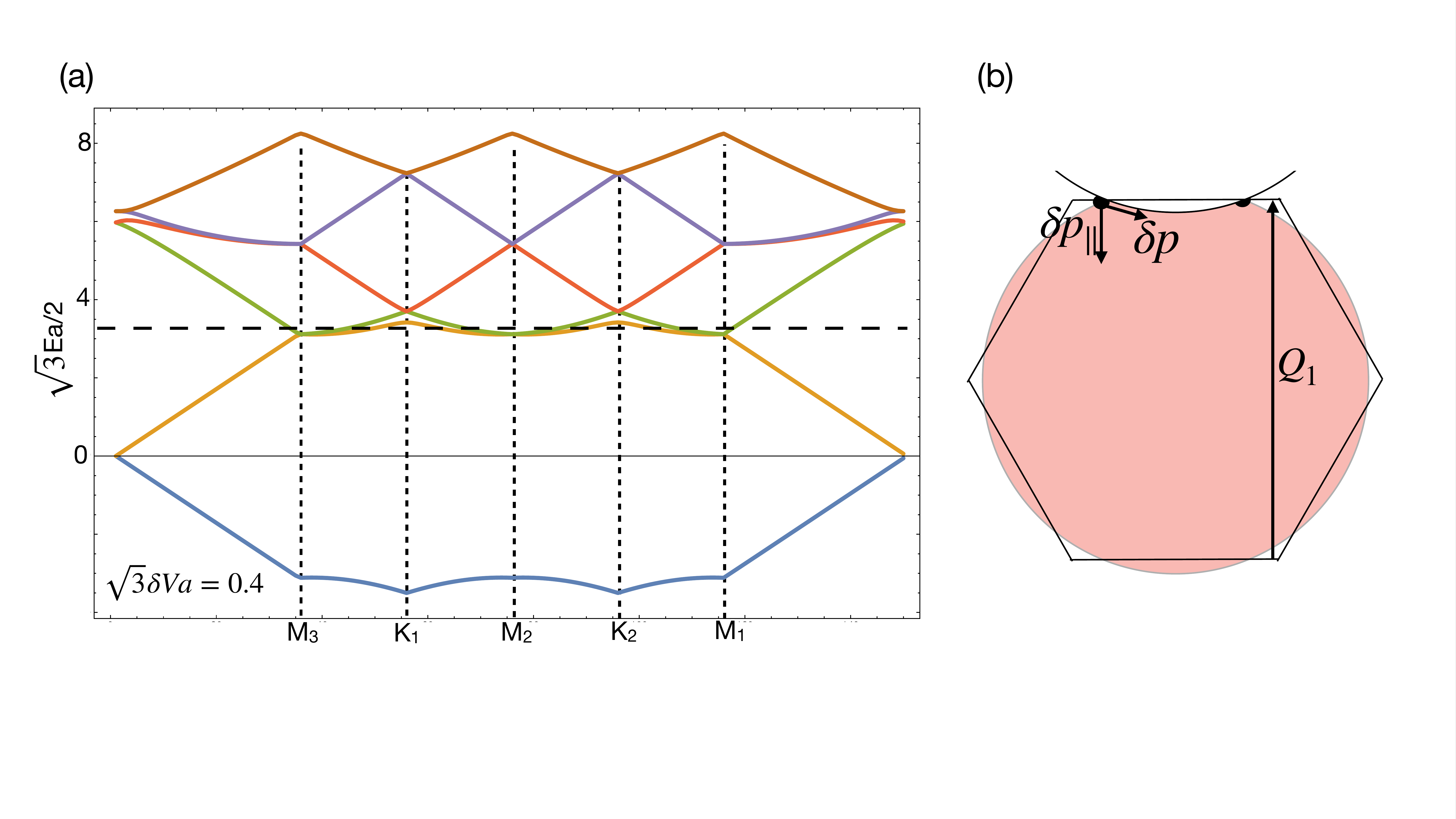}}
    \caption{(a) The dispersion of CFs under exact diagonalization when $B_1$ is small compared with $B_0$. The dashed line indicates Fermi energy that crosses bands, creating CF particle and hole pockets plotted in fig \ref{fig:fs_fqh}(a). (b) An illustration of the integration in eq \eqref{eq:momenta_approx} near a particular intersection point between circular Fermi surface and the Brillouin zone. The red shade denotes momenta $\boldsymbol{p}$'s which scatter to $p-Q_1$ and give nonzero contribution to the integral.}
    \label{fig:supp}
\end{figure}

For a particular $\boldsymbol{p}$ in the first Brillouin zone,  the scattering process involves states with momenta $\boldsymbol{p},\boldsymbol{p}+\boldsymbol{Q}_n,\boldsymbol{p}+2\boldsymbol{Q}_n \cdots$ and we take the momentum cutoff to $\boldsymbol{p}+3\boldsymbol{Q}_n$ for the numerical diagonalization. Fig \ref{fig:supp}(a) shows a dispersion for small $B_1$ which for CFs deforms the circular Fermi surface into particle/hole pockets. 

When solving for the mean field of CFs under an internal magnetic field $b^*$ from chemical potential modulation on physical electrons, we take 
\begin{align}
    \overline a_i=\epsilon^{ij}\partial_j \phi(\boldsymbol{x}),\phi(\boldsymbol{x})=\phi_0\sum_{n=1}^6 \cos (\boldsymbol{Q}_n\cdot \boldsymbol{x}).
\end{align}
The vector potential $\overline a$ thus  has nontrivial Fourier component at reciprocal lattice vectors $\overline a_i(Q_k)=i\epsilon^{ij}Q_{k,j}\phi_0$ where $k=1\cdots 6,i,j=x,y$. This connects CFs states with momentum difference of $Q_k$, i.e. resulting in additional terms
\begin{align}
    \mathcal H_a=\sum_{n=1}^6 \overline a_i(Q_k) \overline\psi_{\boldsymbol{p}}\gamma^i \psi_{\boldsymbol{p}+\boldsymbol{Q}_n}\,.
\end{align}

\subsection{Periodic potential of CFs from periodic magnetic field of physical electrons}

In order to perform our mean field analysis, it was necessary to pass to a CF description with fixed chemical potential, rather than fixed density (i.e. pass from the canonical to the grand canonical ensemble). Here we 
describe how to calculate the profile of the needed CF chemical potential, $V_{\mathrm{CF}}(\boldsymbol{x})$ in Eq.~\eqref{eq: periodic scalar potential}, 
induced by a periodic modulation in the magnetic field. 
From dimensional analysis, it is natural to guess the leading behavior, 
\begin{align}
\label{eq: proportionality V}
|V_1|\sim \frac{|B_1|}{\sqrt{|B_0|}}+\dots\,,
\end{align}
where $V_1$ is defined in Eq.~\eqref{eq: periodic scalar potential} as the amplitude of the oscillating part of the CF chemical potential,
\begin{align}
V_{\mathrm{CF}}(\boldsymbol{x})=
\mu_0+V_1\sum_{n=1,\dots, 6} e^{i \boldsymbol{Q}_n\cdot \boldsymbol{x}}\,.
\end{align}
Indeed, Eq.~\eqref{eq: proportionality V} makes physical sense -- the quantity has dimensions of energy and vanishes as $B_1\rightarrow 0$ (the periodic potential turns off) or $B_0\rightarrow\infty$ (the periodic modulation is small on the scale of the Fermi wave vector).

The proportionality coefficient in Eq.~\eqref{eq: proportionality V} can be determined by enforcing self-consistently,
\begin{align}
  \langle\psi^\dagger \psi(\boldsymbol{x})\rangle_{V_{\mathrm{CF}}(\boldsymbol{x})}=\frac{B(\boldsymbol{x})}{4\pi}\,,
  \end{align}
working power-by-power in $B_1/B_0$. Wick rotating to imaginary time, $t=-i\tau$, 
we single out the $\boldsymbol{Q}_n$ Fourier components of the density one-point function, 
\begin{align}
    \langle\psi^\dagger\psi(\boldsymbol{x})\rangle&=\frac{1}{2\pi}\int d^2\boldsymbol{x}\, \Tr[G_{\psi^\dagger\psi}(\boldsymbol{x},\boldsymbol{x},\tau\rightarrow 0)]e^{i \boldsymbol{Q}_n\cdot \boldsymbol{x}}=\frac{B(\boldsymbol{Q}_n)}{4\pi}\,,\qquad
    G_{\psi^\dagger\psi}(\boldsymbol{x},\boldsymbol{x},i\omega)= (i\omega+i\gamma_0\gamma^i \partial_i+V_{\mathrm{CF}}(\boldsymbol{x}))^{-1}.
    \label{eq:mu_osci}
\end{align}
In the regime where $|V_1|\ll |\mu_0|$, i.e. $B_1\ll B_0$, we can expand Eq.~\eqref{eq:mu_osci} with a small parameter $\frac{V_1}{\mu_0}$,
\begin{align}
G_{\psi^\dagger\psi}(\boldsymbol{x},i\omega)=G_0+\delta G=G_0-G_0 (V_{\mathrm{CF}}(\boldsymbol{x})-\mu_0) G_0+\dots\,, \qquad G_0=(i\omega+i\gamma_0\gamma^i\partial_i+\mu_0)^{-1}\,.
\end{align}
Keeping only the leading (one-loop) contribution, we obtain for the oscillating part, $\delta G$,
\begin{align}
    \delta G=-\int d^2 \boldsymbol{p} d\omega \Tr[G_0(\boldsymbol{p},i\omega)V_1 G_0(\boldsymbol{p}+\boldsymbol{Q}_n,i\omega) ]=\frac{B_1(\boldsymbol{Q}_n)}{4\pi}.
    \label{eq:mu_fourier}
\end{align}

We find it convenient to express the Green's function in the eigen-energy basis for each momentum $\boldsymbol{p}$, rewriting $G_0(\boldsymbol{p},i\omega)=P(\boldsymbol{p}) (i\omega+|p|+\mu_0)^{-1}+(1-P(\boldsymbol{p}))(i\omega-|p|+\mu_0)^{-1}$ in terms of the projection operator, $P(\boldsymbol{p})$, 
that project the Dirac spinor onto the positive energy band. 
The frequency integral has the general form,
\begin{align}
    \int d\omega\, (i\omega+\epsilon_1)^{-1}(i\omega+\epsilon_2)^{-1}=1-\Theta(-\epsilon_1\epsilon_2)\frac{1}{|\epsilon_1-\epsilon_2|}\,.
\end{align}
Here $\Theta(X)$ is the Heaviside step function, which tells us that the negative energy states (well below the chemical potential) do not contribute. 
Furthermore, for states with positive energy, we observe that only when $\boldsymbol{p}$ and $p+\boldsymbol{Q}_n$ are 
above and below the Fermi surface, will the integral in Eq.~\eqref{eq:mu_fourier} be nonzero. The dominant contribution comes when the energies at momenta $\boldsymbol{p},\boldsymbol{p}+\boldsymbol{Q}_n$ are close. For the specific $\boldsymbol{Q}_n$ for the hexagonal Brillouin zone, one  concludes that the integral is most singular at the momenta where the original circular Fermi surface (for integer filling, i.e. with the same area as the first Brillouin zone) intersects the hexagonal BZ. Linearizing the deviation from one intersection point as $\delta \boldsymbol{p}$, we have for the integral in Eq.~\eqref{eq:mu_fourier} (see Fig.~\ref{fig:supp}(b))
\begin{align}
\int d(\delta p)d(\delta p_\parallel)  \frac{\xi_{\boldsymbol{p},\boldsymbol{p}+\boldsymbol{Q}_n}}{v_\perp \delta p+v_\parallel\delta p_\parallel}V_1=\frac{B_1(\boldsymbol{Q}_n)}{4\pi},
\label{eq:momenta_approx}
\end{align}
where $\xi_{\boldsymbol{p},\boldsymbol{p}+\boldsymbol{Q}_n}=\Tr [P(p)P(p+\boldsymbol{Q}_n)]$ and $\delta p_\parallel$ is the momentum projected along $\boldsymbol{Q}_n$ (not to be confused with the momentum along the Fermi surface).  
The above integral is regular and can be calculated numerically. However, since the regime of interest corresponds to the case $B_1/B_0,V_1/V_0\sim\mathcal{O}(1)$, we do not expect the final coefficient, $C\sim\mathcal{O}(1)$ to be physically meaningful. We therefore 
confirm the expectation from dimensional analysis and estimate,
\begin{align}
    C\sqrt {B_0}\, V_1=\frac{B_1(\boldsymbol{Q}_n)}{4\pi}\,
   \end{align}
   where the factor $\sqrt{B_0}$ comes in from the $|\boldsymbol{Q}_n|$ dependence on the unit cell size and hence $B_0$, given that unit cell size $S$ satisfies $B_0 S=4\pi$. We have $|\boldsymbol{Q}_n|\propto \sqrt {B_0}$ and the expression \eqref{eq:momenta_approx} scales as $|\boldsymbol{Q}_n|\sim \sqrt{B_0}$. 

   
\subsection{Position of the emergent CF Dirac cones}

\label{dirac_symmetry}
We show that the emergent Dirac cones for CFs  are pinned at certain high-symmetry points - including $M_i,K(K')$ at BZ edge and $\Gamma$ at BZ center. 

Two threads of arguments are presented: From the anomaly of the CFs, i.e. parity anomaly in $\mathcal {PH}\rtimes U(1)_{\psi}$, at mean-field level for a fixed energy, the spectrum has to be either gapless or possess odd number of Dirac cones to match the anomaly. For the Dirac cone case, note that $\mathcal {PH}$ sends momenta $k\rightarrow -k$ due to its anti-unitary action. Hence generically the Dirac cones come in pairs, except at high symmetry point $M_i$'s since $\pm M_i$ differ by a reciprocal lattice vector and one could have odd number of Dirac cones pinned at $M_i$'s. Another possibility is a single Dirac cone at $\Gamma$, which is $\mathcal {PH}$ invariant. For the filling $1$ for CFs focused in the main text, we have the case for Dirac cones emerging at $M_i$'s.

Another argument follows from symmetry of the CF action: $\mathcal {PH}$ leaves $M_i,\Gamma$'s invariant and squares to $-1$. Hence there is a Kramer's degeneracy at $M_i,\Gamma$'s.

To see the appearance of Dirac cones at $K,\tilde K=-K$ points, which in our case occurs between the second and third Bloch bands in energy (along with the Dirac cone at $\Gamma$ to cancel the parity anomaly), we consider two relevant symmetries $C_3, \mathcal {PH}\cdot\mathcal {P}$, with the action on the CFs
\begin{align}
   C_3: \psi(r)\rightarrow e^{-i\sigma^z\frac{\pi}{3}} \psi(C_3(r)),\nonumber\\
   \mathcal {PH}\cdot \mathcal P: \psi\rightarrow \sigma^z \psi.
   \end{align}
   We show that acting on the eigenstates with momenta $K$ and its $C_3$ equivalents $K',K''$, the two symmetries do not commute, which implies degeneracy. The Bloch wavefunctions with positive energy for the  CFL action eq \eqref{eq:cfaction} are taken to be
   \begin{align}
   \psi_K:(1,e^{i\phi})^T,\psi_{K'}:C_3(\psi_K)=(e^{-2i\phi},e^{i3\phi})^T,\nonumber\\
   \psi_{K''}:-C_3^{-1}(\psi_K)=(e^{-4i\phi},e^{i5\phi})^T.
   \end{align}
   Under such basis, the transformation matrix for $C_3, \mathcal {PH\cdot P}$ reads
   \begin{align}
   C_3: \left (\begin{array}{ccc} 0&1&0\\ 0&0&1\\1&0&0\end{array}\right ),\nonumber\\
   \mathcal{PH\cdot P}: \left (\begin{array}{ccc}  0&e^{2i\phi}&0\\ e^{2i\phi}&0&0\\0&0&-e^{2i\phi}\end{array}\right ).
   \end{align}
   which do not commute. Hence the degeneracy at $K,\tilde K=-K$ are enforced.

\subsection{Symmetry action on the emergent Dirac cones}


Here we specify how different symmetries act on the emergent Dirac fermions, $\chi_I$, $I=1,2,3$, of the emergent QED$_3$ theory. We start with the lattice translations and $C_3$ rotations,
\begin{align}
  T_i:\chi_I&\rightarrow e^{i\boldsymbol{M}_I\cdot \boldsymbol{x}}\,\chi_I,\nonumber\\
  C_3:\chi_I&\rightarrow e^{i\gamma^t\frac{\pi}{3}}\,\chi_{(I+1)\operatorname{mod}3},
  \end{align}
Here $\boldsymbol{M}_I$ are the reciprocal lattice vectors corresponding to the three $M$-points. As described in the main text, at long wavelengths the $C_3$ rotation symmetry is enhanced to an emergent SU$(3)$ flavor symmetry. 

  The discrete symmetries of most interest to us are the anti-unitary  (electronic) particle-hole and parity symmetries, which we denote $\mathcal{CT}$ (the product of charge conjugation and time-reversal) and $\mathcal{CP}$,
  \begin{align}
  \mathcal {CT}:\chi_I&\rightarrow -i\tau^y \chi_I(-t,\boldsymbol{x})\,,\qquad(a_0,a_i)\rightarrow (a_0,-a_i)\nonumber\\
  \mathcal {CP}:\chi_I&\rightarrow \tau^x \chi_{s(I)}(t,-x,y)\,, \qquad (a_0,a_i)\rightarrow (a_0,-a_x,a_y)\,,
\end{align}
where $s(I)$ permutes the Dirac fermion species as $s(1)=1,s(I=2,3)=5-I$. The emergent $N_f=3$ QED$_3$ theory is invariant under these symmetries. Both are broken by the introduction of a periodic scalar potential, i.e. the Dirac mass operator.




\subsection{Topological orders of the FCI states}

Here we consider the topological orders of the fractional Chern insulator states considered in the main text. This requires introducing an auxiliary gauge field, $b_\mu$, to the $N_f=3$ QED$_3$ theory such that all Chern-Simons terms are properly quantized and the theory is gauge invariant,
\begin{align}
S=\int_{t,\boldsymbol{x}}\sum_{I=1}^3i\bar\chi_I(\partial_\mu-ia_\mu)\gamma^\mu\chi_I-\frac{1}{8\pi}ada-\frac{2}{4\pi}bdb+\frac{1}{2\pi}bd(a+A')-\frac{1}{8\pi}A'dA'\,.
\end{align}
Note here that $a_\mu$ is a $\mathrm{spin}_c$ connection, while $b_\mu$ is an ordinary U$(1)$ connection. See e.g. Refs.~\cite{Seiberg2016,Goldman2020fermionization} for a more detailed discussion of the difference between $\mathrm{spin}_c$ and U$(1)$ connections. For our needs, the only important consequence of this distinction is that anyons associated with $a_\mu$ will have their statistics shifted by $\pi$, since spin$_c$ connections couple to fermions. Note also that we continue to consider the theory on the surface of a TI, the bulk of which gives rise to the final $A'dA'/8\pi$ term.

One can easily recover the action discussed in the main text,  Eq.~\eqref{eq: final effective action}, by integrating out $b_\mu$, meaning that the local physics of the two theories is equivalent. However, unlike Eq.~\eqref{eq: final effective action}, this theory is gauge invariant on any manifold. This is crucial for correctly diagnosing the topological order on integrating out the $\chi_I$ fields deep in a FCI phase. 

We begin by considering the effect of a singlet mass term, $\mathcal{L}_{\mathrm{mass}}=-m\sum_I\bar\chi_I\chi_I$, which generated by the periodic scalar potential in Eq.~\eqref{eq: periodic mu}. This deformation causes the $\chi_I$ fermions to form Chern insulators, each with Chern number $\sgn(m)/2$. The resulting state is a FCI described by the topological quantum field theory (TQFT) with Lagrangian,
\begin{align}
\label{eq: mass TQFT}
\mathcal{L}_{\mathrm{FCI}}&=\frac{3\sgn(m)-1}{2}\frac{1}{4\pi}ada-\frac{2}{4\pi}bdb+\frac{1}{2\pi}bd(a+A')-\frac{1}{8\pi}A'dA'\,.
\end{align}
The Hall conductivity of this state can be calculated by integrating out both gauge fields,
\begin{align}
\mathcal{L}_{\mathrm{response}}&=\left(\frac{1}{2}-\frac{\sgn(m)}{6}\right)\frac{1}{4\pi}A'dA'-\frac{1}{8\pi}A'dA'\,,
\end{align}
where we have separated out the bulk TI contribution. Indeed, the first term has coefficient $1/3$ ($m>0$) or $2/3$ ($m<0$), suggesting that these states are have the topological order of the $\nu=\pm 1/3$ Laughlin states. We can furthermore calculate the ground state degeneracy on the torus, which is $|\det K|=3$ (where $K$ is the $(3\sgn(m)/2-1/2,-2,1)$ $K$-matrix corresponding to Eq.~\eqref{eq: mass TQFT}, in agreement with this expectation. 

The equivalence can be established more concretely at the level of the TQFT. We start with the case $m>0$. In this case, the first term in Eq.~\eqref{eq: mass TQFT} is a trivial TQFT, $ada/4\pi$. Indeed, in this case we can integrate out $a$ in a gauge invariant way to obtain,
\begin{align}
\mathcal{L}_{m>0}=-\frac{3}{4\pi}bdb+\frac{1}{2\pi}bdA'-\frac{1}{8\pi}A'dA'\,.
\end{align}
This is of course the usual Laughlin U$(1)_{-3}$ state, with the same charge-$1/3$ anyonic quasiparticles as the usual Laughlin FQH state. Notice that the fluctuating gauge field here is an ordinary U$(1)$ connection, as the spin$_c$ connection has been eliminated.

We now consider the case of $m<0$. Here we cannot immediately integrate out $a_\mu$, but we can nevertheless argue its equivalence to a Laughlin state. We start by using a trick to simplify the TQFT (which in this case incorporates the same physics as the more rigorous notion of level-rank duality; see Refs.~\cite{Goldman2018a,Goldman2020fermionization} for an introduction accessible to condensed matter physicists). We start by introducing a background composite fermion world-line, $j^\mu_{f}$, coupling to $a_\mu$. Because $j^\mu_f$ describes a fermion, we can use flux attachment to rewrite it in terms of a bosonic world-line variable, $J^\mu$, with a single flux attached,
\begin{align}
j_f^\mu a_\mu\rightarrow J^\mu\,\beta_\mu+\frac{1}{4\pi}(\beta+a) d(\beta+a)\,,
\end{align}
where $\beta_\mu$ is a new U$(1)$ gauge field implementing the flux attachment. Its role is to explicitly implement the aforementioned shift of the anyon statistics by $\pi$ associated with the fact that $a_\mu$ is a spin$_c$ connection. Therefore, rather than working with a TQFT where one probes with fermionic lines, we can equivalently work with a different representation of the TQFT where all of the probes are bosonic. We therefore arrive at an equivalent TQFT,
\begin{align}
\mathcal{L}_{m<0}&=\frac{1}{4\pi}\beta d\beta+\frac{1}{2\pi}\beta da-\frac{1}{4\pi} ada-\frac{2}{4\pi}bdb+\frac{1}{2\pi}bd(a+A')-\frac{1}{8\pi}A'dA'\,,
\end{align}
where we have dropped the dependence on the probe, $J^\mu$. Now again the theory involves a trivial TQFT for $a_\mu$, and we can integrate it out in the same way as before to obtain,
\begin{align}
\mathcal{L}_{m<0}&=\frac{2}{4\pi}\beta d\beta+\frac{1}{2\pi}bd(\beta+A')-\frac{1}{4\pi}bdb-\frac{1}{8\pi}A'dA'\,,
\end{align}
which, incidentally, is the TQFT we would have found had we invoked the level-rank duality between U$(1)_{-2}$ (with a spin$_c$ connection) and U$(1)_2$ (with an ordinary U$(1)$ connection). We can integrate out $b_\mu$ to obtain the properly quantized, single-component TQFT,
\begin{align}
\mathcal{L}_{m<0}&=\frac{3}{4\pi}\beta d\beta+\frac{1}{2\pi}\beta dA'+\frac{1}{8\pi}A'dA'\,.
\end{align}
From here, one can easily confirm that this theory has the correct Hall conductivity and ground state degeneracy. We therefore observe that the $m<0$ state is equivalent to a U$(1)_{3}$ TQFT, which is the particle-hole conjugate of the U$(1)_{-3}$ state found for $m>0$.

In the case of the FCI sequence accessed by filling $q\in\mathbb{Z}$ Landau levels of the $\chi_I$ fermions, the TQFT one obtains is
\begin{align}
\mathcal{L}_{\textrm{FCI Jain}}&=\frac{3q-2}{4\pi}ada-\frac{2}{4\pi}bdb+\frac{1}{2\pi}bd(a+A')-\frac{1}{8\pi}A'dA'\,.
\end{align}
Notice that the $q=0$ and $q=1$ states are the same as the states we found by turning on mass operators with $m<0$ and $m>0$ respectively. This general sequence of states have $K$-matrices that each correspond to a state on the principal Jain sequence. Indeed, each Jain state can be described by a $2\times 2$ $K$-matrix with label $(k,-2,1)$, $k\in\mathbb{Z}$, where the gauge field with level $k$ is a spin$_c$ connection (i.e. couples to composite fermions). The anyon content of each of these states can be found discussed in numerous references, e.g. Refs.~\cite{Lopez-1991,Lopez1999}.

\end{document}